\documentclass[prb,twocolumn]{revtex4}
\usepackage{graphicx}
\usepackage{hyperref}
\usepackage{amssymb}
\usepackage{dcolumn}
\usepackage{float}
\usepackage{bm}
\usepackage{multirow}
\usepackage{booktabs}
\usepackage{amsmath} 
\usepackage[utf8]{inputenc}
\usepackage{lmodern}
\usepackage{color}
\usepackage{makecell}

\newcolumntype{M}[1]{>{\centering\arraybackslash}m{#1}}

\newcommand{\bs}{\boldsymbol}
\DeclareMathAlphabet{\bi}{OML}{cmm}{b}{it}

\def\be{\begin{equation}}
\def\ee{\end{equation}}
\def\bearr{\begin{eqnarray}}
\def\eearr{\end{eqnarray}}

\usepackage{tikz}
\usepackage{pgfplots}

\begin{document}
\title{Effect of electron-hole asymmetry on optical conductivity in 8-$Pmmn$ borophene} 
\bigskip
\author{Sonu Verma, Alestin Mawrie and Tarun Kanti Ghosh\\
\normalsize
Department of Physics, Indian Institute of Technology-Kanpur,
Kanpur-208 016, India}
\begin{abstract}
We present a detail theoretical study of the Drude weight and 
optical conductivity of 8-$Pmmn$ borophene having tilted anisotropic 
Dirac cones. 
We provide exact analytical expressions of $xx$ and $yy$ components of 
the Drude weight as well as maximum optical conductivity. We also obtain 
exact analytical expressions of 
the minimum energy ($\epsilon_1$) required to trigger the optical transitions 
and energy ($\epsilon_2$) needed to attain maximum optical conductivity. 
We find that the Drude weight and optical conductivity are highly anisotropic as a 
consequence of the tilted Dirac cone. The tilted parameter can be extracted by 
knowing $\epsilon_1$ and  $\epsilon_2$ from optical measurements. The maximum 
values of the components of the optical conductivity do not depend on the carrier 
density and the tilted parameter. The product of the maximum values of the anisotropic 
conductivities has the universal value $(e^2/4\hbar)^2$.
The tilted anisotropic Dirac cones in 8-$Pmmn$ borophene can be realized by 
the optical conductivity measurement.
\end{abstract}

\maketitle

\section{introduction}
Graphene is the first atomically thin two-dimensional (2D) material 
having isotropic Dirac cones realized in a laboratory 
\cite{graphene1,graphene2}. 
Since then, there have been numerous attempts to synthesize 
more and more new 2D materials having Dirac cones.
Several quasi-2D materials possessing Dirac cones such as 
silicene\cite{silicene}, germanene\cite{germanene},
and MoS${}_2$ \cite{mos2} have been synthesized experimentally and are being studied
theoretically.

Recently, there has been intense research interest in
synthesis of 2D crystalline boron structures, referred to as borophene.
Several attempts have been made to synthesize a stable structure of
borophene but only three different quasi-2D structures of borophene have 
been synthesized \cite{boron-syn}. Various numerical experiments have 
predicted a large number of borophene structures with various geometries 
and symmetries \cite{boron0,boron1}. The orthorhombic 8-$Pmmn$ borophene is 
one of the energetically stable structures, having ground state energy lower 
than that of the $\alpha$-sheet structures and its analogues.
The $Pmmn$ boron structures have two non-equivalent sub-lattices.
The coupling and buckling between two sub-lattices and vacancy give rise 
to the energetic stability as well as tilted anisotropic Dirac cones \cite{boron2}.
The coupling between different sub-lattices enhances the strength of the
boron-boron bonds and hence gives rise to structural stability.
The finite thickness is required for energetic stability of 2D
boron allotropes. 
The orthorhombic 8-$Pmmn$ borophene possesses tilted anisotropic Dirac cones and
is a zero-gap semiconductor. It can be thought of as topologically 
equivalent to the distorted graphene.

In the last couple of years, there have been several theoretical 
studies on 8-$Pmmn$ borophene.
Very recently, electronic properties of 8-$Pmmn$ borophene have been studied using the 
first-principle calculations and have shown the Dirac cones arising from the $p_z$ 
orbitals of one of the two inequivalent sub-lattices \cite{boron2}.
Zabolotskiy and Lozovik proposed a tight-binding Hamiltonian 
for 8-$Pmmn$ borophene and obtained a low-energy effective Hamiltonian in
the continuum limit \cite{boron3}. The effective Hamiltonian successfully described
all the main features of the quasi-particle spectrum as predicted in {\it ab initio} 
calculation.
A similar Hamiltonian has been considered in Ref. \cite{hamil} for studying 
quinoid-type graphene due to mechanical deformation and 
organic compound $\alpha$-(BEDT-TTF)${}_2$I${}_3$.
The anisotropic plasmon dispersion of borophene is predicted in Ref. \cite{amit}. 
The magnetotransport coefficients have been investigated very recently \cite{firoz}.

The frequency-dependent optical conductivity is associated with the transitions 
from a filled band to an empty band, whereas the zero-frequency Drude weight is 
due to the intra-band transitions. The real part of the complex optical
conductivity is connected with the absorption of the incident photon energy.
Its measurement is an important tool for 
extracting the shape and nature of the energy bands. 
There are several theoretical and experimental studies on the optical conductivity
in various monolayer quantum materials such as graphene 
\cite{graphene-op-con,graphene-op-con1,op-con-exp}, 
silicene \cite{silicene-op-con,silicene-op-con1,silicene-op-con2}, 
MoS${}_{2}$ \cite{mos2-op-con,mos2-op-con1}, and 
surface states of topological insulators \cite{ti-op-con,ti-op-con1,ti-op-con2}.

In this paper, we study zero-frequency Drude weight and frequency-dependent optical 
conductivity of 8-$Pmmn$ borophene. 
We find that the Drude weight and optical conductivity are highly anisotropic due 
to tilted Dirac cones. 
We obtain an analytical expression of the minimum photon energy 
required for triggering the optical transitions and of the photon energy at which the 
conductivities attain maximum value.    
The maximum value of the optical conductivity along the tilted and perpendicular
directions are obtained, which are independent of the carrier density and tilting parameter.
The spectroscopic measurement of the absorptive part of the optical conductivity can shed
some light on the anisotropic but tilted Dirac cone.

This paper is organized as follows. In Sec. II, we provide basic 
information of 8-$Pmmn$ borophene in detail. 
We discuss the Drude weight and absorptive part of the optical conductivity 
in Sec. III. An alternate derivation of the optical conductivity using 
Green's function method is provided in the Appendix.
We provide a summary and conclusions in Sec. IV.

\section{Basic information}
The massless Dirac Hamiltonian associated with the 8-$Pmmn$ borophene 
in the vicinity of one of the two independent Dirac points is given 
by \cite{boron3}                                                                                  
\begin{eqnarray} \label{Hamil} 
H = v_{x} \sigma_{x} p_{x}+ v_{y}\sigma_{y}p_{y}+ v_{t}\sigma_{0}p_{y},
\end{eqnarray}
where $p_\mu$ with $\mu=x,y$ are the momentum operators, $\sigma_\mu $ are 
the $2\times 2$ Pauli matrices, and $\sigma_0 $ is the $2 \times 2 $ identity matrix. 
The velocities are given \cite{boron3} as $v_{x} = 0.86 v_F$, $v_{y}= 0.69 v_F$, 
and $v_{t}= 0.32 v_F$ with $v_F = 10^6$ ms$^{-1}$. The Hamiltonian associated with the
second Dirac cone has the opposite sign of $v_t$.

The energy dispersion and the corresponding wave functions are given by
\begin{eqnarray}
E_{\lambda}({\bf k}) = \hbar k  [v_{t} \sin{\theta_{\bf k}} + \lambda \Delta(\theta_{\bf k})]
\end{eqnarray}
and
\begin{eqnarray}
\psi_{\bf k}^{\lambda}({\bf r}) = \frac{e^{i\textbf{k} \cdot \textbf{r}}}{\sqrt{2}}
\begin{pmatrix}
1\\
\lambda e^{i \phi}
\end{pmatrix},
\end{eqnarray}
where $\lambda = \pm $ denotes the conduction and valence bands, respectively,
$\theta_{\bf k} = \tan^{-1}(k_y/k_x)$, 
$\Delta(\theta_{\bf k}) = \sqrt{v_{x}^{2}\cos^{2}{\theta_{\bf k}} + 
v_{y}^{2}\sin^{2}{\theta_{\bf k} }}$
describes anisotropy of the spectrum and $ \phi = \tan^{-1}(v_y k_y/v_x k_x)$.
The energy difference between the conduction and valence bands at a given ${\bf k}$
is $E_g({\bf k}) = E_+({\bf k}) - E_{-}({\bf k})  =2 \hbar k \Delta(\theta_{\bf k}) $.
Note that the first term in the energy spectrum tilts the Dirac cone and breaks the
electron-hole symmetry $E_{\lambda}({\bf k}) = - E_{-\lambda}({\bf k})$, 
even for the isotropic case $ v_x = v_y$. The tilted Dirac cones are depicted in Fig. 1. 

\begin{figure}[htbp]
\begin{center}\leavevmode
\includegraphics[width=65mm,height=45mm]{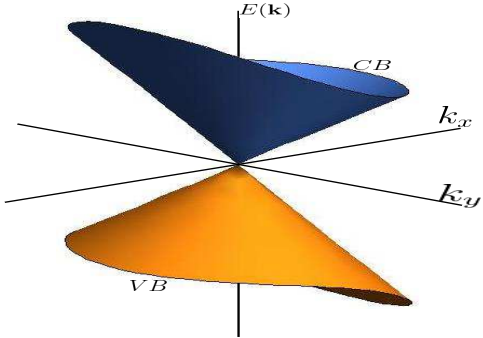}
\caption{Plots of $E$-${\bf k}$ dispersion displaying the tilted anisotropic Dirac cones.}
\label{Fig1}
\end{center}
\end{figure}

The Berry connection of 8-$Pmmn$ borophene is given by
\begin{eqnarray}
{\bf A}_{\lambda}({\bf k}) = - \frac{v_x v_y}{2 \Delta^2(\theta_{\bf k})} 
\, \frac{\hat{\theta}_{\bf k} }{k},
\end{eqnarray}
where $\hat{\theta}_{\bf k} = - \sin\theta_{\bf k} \, \hat{x} + \cos\theta_{\bf k} \, \hat{y}$ 
is the unit polar angle. 
The corresponding Berry phase is $\gamma_{\lambda} = 
\oint {\bf A}_{\lambda}({\bf k}) \cdot d{\bf k}= \pi$, exactly the same as in the 
monolayer graphene case.

The chirality operator can be defined as
\begin{eqnarray}
{\hat \Lambda} = \frac{v_x \cos\theta_{\bf k} \, \sigma_x + v_y\sin\theta_{\bf k} \, \sigma_y}{\Delta(\theta_{\bf k})}.
\end{eqnarray}
It can be easily checked that the chirality operator $ {\hat \Lambda}$ commutes with 
the Hamiltonian $H$ even in the presence of the tilted Dirac cone and 
the eigenfunctions $\psi_{\bf k}^{\lambda}({\bf r}) $ are also eigenfunctions of the chirality 
operator with the eigenvalues $\lambda = \pm 1$, respectively.

The role of the Berry phase $ \gamma_{\lambda} = \pi $ and of the
chiral symmetry preservation  must be reflected in the scattering process. This can be
easily understood by analyzing the angular scattering probability for the borophene. 
This is given by the squared moduli of the overlap matrix element between the initial spinor 
($\chi_{\lambda}(\theta_{\bf k}) $) and the final spinor ($\chi_{\lambda}(\theta_{\bf k}^{\prime}) $)
with $| {\bf k}  | = |{\bf k}^{\prime} |$.
The angular scattering probability is then 
\begin{eqnarray}
|f(\theta_{\bf k},\theta_{{\bf k}^{\prime}})|^2 & = &
|\langle \chi_{\lambda}(\theta_{\bf k}) 
|\chi_{\lambda}(\theta_{{\bf k}^{\prime})} \rangle |^2 \nonumber \\
& = & \frac{1}{2} \Big[1 + \frac{v_{x}^{2} 
\cos \theta_{\bf k} \cos \theta_{{\bf k}^{\prime}} +
v_{y}^{2}\sin \theta_{\bf k} \sin \theta_{{\bf k}^{\prime}}}
{\Delta(\theta_{\bf k})\Delta(\theta_{{\bf k}^{\prime}})}\Big] \nonumber.
\end{eqnarray}
It is to be noted that the wave functions do not depend on the tilt parameter $v_t$.
Hence the Berry connection and $|f(\theta_{\bf k},\theta_{{\bf k}^{\prime}})|^2 $ are
also independent of the tilt parameter.
It also shows that $ |f(\theta_{\bf k},\theta_{{\bf k}^{\prime}})|^2$ is independent of the bands
and vanishes exactly when 
$\theta_{{\bf k}^{\prime}} - \theta_{\bf k} = \pi$. It implies that the 
backscattering is completely absent, similar to the graphene case.
The absence of backscattering survives even for tilted anisotropic energy spectrum.
This is due to the conservation of the chirality and/or the $\mp \pi $ Berry phase.


The density of states is given by
\begin{eqnarray}
D(E) & = & g_s g_v \int \frac{d^{2}k}{(2 \pi)^2} 
\delta(E - E_{\lambda}({\bf k})) \nonumber \\ 
& = & N_{0} \frac{|E|}{\pi^{2}\hbar^{2}v_F^2},
\end{eqnarray}
where the spin degeneracy $g_s =2 $ and the ``valley degeneracy" $g_v =2 $ \cite{lozovik1}. 
Also, the constant $N_{0} $ is given by
$$ N_{0} = 
\int_{0}^{2\pi} \frac{ v_F^2  \; d \theta_{\bf k}}{[v_t \sin \theta_{\bf k} + 
\lambda \Delta(\theta_{\bf k})]^{2}} = 15.2263. 
$$ 
For a given carrier density $n_c$, the Fermi energy is 
$ E_F =  \hbar \tilde v_{F} \sqrt{\pi n_c} $ with $ \tilde v_F = \sqrt{2\pi/N_0} v_F$
and the associated anisotropic Fermi wave vectors are  
obtained as 
\begin{eqnarray}
k_F^{\lambda} (\theta_{\bf k}) & = & 
\frac{E_F }{ \hbar |v_t \sin \theta_{\bf k} + \lambda \Delta(\theta_{\bf k})|}.
\end{eqnarray}

The components of the velocity operator along the $x$- and
$y$-directions are 
$\hat{v}_x = v_x \sigma_x$ and $\hat{v}_y = v_t \sigma_0 + v_y \sigma_y$.
The expectation values of these operators are given by
$\langle \hat{v}_{x}\rangle_{\lambda} = [ v_{x}^2/\Delta(\theta_{\bf k})] \cos{\theta_{\bf k} }$ 
and $\langle \hat{v}_{y}\rangle_{\lambda} = v_{t} + [v_{y}^2/\Delta(\theta_{\bf k})] \sin{\theta_{\bf k} }$, 
respectively.

\section{optical conductivity}
We consider $n$-doped 8-$Pmmn$ borophene subjected to zero-momentum 
electric field ${\bf E} \sim \hat {\bs \mu} E_0 e^{i \omega t} $ with
oscillation frequency $\omega$ ($ \hat {\bs \mu} = \hat {\bf x}, \hat {\bf y} $).
The complex charge optical conductivity tensor is given by 
$\Sigma_{\mu \nu}(\omega) = \delta_{\mu\nu} \sigma_{D}(\omega) + 
\sigma_{\mu\nu}(\omega)$,
where $\mu,\nu=x,y$, $\sigma_{D}(\omega) = \sigma_d/(1- i \omega \tau)$ is 
the dynamic Drude conductivity due to the intra-band transitions, with 
$\sigma_d$ being the static Drude conductivity and 
$\sigma_{\mu \nu}(\omega) $ being the complex optical conductivity due to
transitions between valence and conduction bands.
It should be mentioned here that Re $\sigma_D$ and Re $\sigma_{\mu \nu} $
correspond to the absorption of the photon energy.

{\bf Drude weight}: 
The Drude weight at vanishingly low-temperature is given by \cite{mermin}
\begin{eqnarray}
D_{\mu\nu}^{} = \frac{g_s g_v e^2}{4\pi}\int d^2 k \,
\langle\hat{v}_{\mu}\rangle_{} \langle\hat{v}_{\nu}\rangle_{} 
\delta(E({\bf k})-E_{F}^{} ),
\end{eqnarray}

On further simplification, we obtain
\begin{eqnarray}
D_{\mu \nu}^{} = \frac{e^2}{\hbar}\frac{E_{F} }{\pi\hbar} 
\delta_{\mu\nu}(\delta_{\mu x} N_1^{} + \delta_{\nu y} N_2^{}),
\end{eqnarray}
where $ N_{1} = 4.686 $ and $ N_2 = 2.673$. 
In this case, the Drude weight is anisotropic, unlike the monolayer graphene case 
where the Drude weight $D_{w}^{G} = (v_F e^2/\hbar) \sqrt{\pi n_c}$ is isotropic.

{\bf Optical Conductivity}: Within the linear response theory, 
the Kubo formula for the optical conductivity tensor $\sigma_{\mu\nu}(\omega)$ 
is given by
\begin{eqnarray}
\sigma_{\mu\nu}(\omega) & = & \frac{1}{\hbar(\omega+i \eta)}\int_{0}^{\infty}dt 
e^{i(\omega+i\eta)t} \langle [\hat j_{\mu}(t), \hat j_{\nu}(0) ] \rangle, 
\nonumber
\end{eqnarray}
where 
$ \langle [\hat j_{\mu}(t), \hat j_{\nu}(0) ]\rangle = 
\sum_{m,n}[f(E_n)-f(E_m)] e^{i(E_n-E_m)t/\hbar}j_{\mu}^{nm}j_{\nu}^{mn}$, 
$\hat j_{\mu} = e \hat v_{\mu} $ is the charge current density with $\mu = x,y$, 
$f(E)$ is the Fermi-Dirac 
distribution function and $\eta \rightarrow 0^{+}$. 
Here $E_n$ and $E_m$ are the discrete energy levels of the system. 
So changing the sum into integration over momentum space, the real part of 
the charge optical conductivity is given by
\begin{eqnarray} \label{opt}
{\rm Re\;\sigma_{\mu\nu}(\omega)} & = & 
\frac{e^2}{4\pi \omega}\int d^2k [f(E_{-}(\textbf{k})) - f(E_{+}(\textbf{k}))] \nonumber \\ 
& & v_{\mu}^{-+}({\bf k})  v_{\nu}^{+-}({\bf k}) 
\delta(E_g({\bf k}) - \hbar \omega).
\end{eqnarray}
The final expression of the real part of the optical conductivity tensor is given by
\begin{widetext}
\begin{eqnarray} \label{opt1}
{\rm Re \; \sigma_{\mu\nu}}(\omega)  
 =  \frac{e^2}{4 \pi \hbar}\int_{0}^{2\pi}  d\theta_{\bf k} 
\frac{v_{x}^2 v_{y}^2}{ \Delta^4(\theta_{\bf k} )} 
[ f(E_-) - f(E_+) ]  
[(\delta_{\mu x}\sin^2{\theta_{\bf k}} + \delta_{\nu y}\cos^2{\theta_{\bf k} })  
\delta_{\mu\nu}-(1-\delta_{\mu\nu})\sin{\theta_{\bf k} }\cos{\theta_{\bf k}} ],
\end{eqnarray}
\end{widetext}
where 
$ E_{\pm}(k_{\omega}(\theta_{\bf k}),\theta_{\bf k}) \equiv E_{\pm} $ with
$k_{\omega}(\theta_{\bf k}) = \omega/[2 \Delta(\theta_{\bf k})]$.

First of all, we find that the real part of the off-diagonal optical conductivity 
${\rm Re} \; \sigma_{xy}(\omega)$ vanishes exactly.
For monolayer graphene ($v_t=0$ and $v_x=v_y=v_F$), Eq. (\ref{opt1}) gives featureless 
isotropic optical conductivity which has a step-like shape with a step height 
$\sigma_0 = e^2/4\hbar$ at $\hbar \omega = 2 E_F^0 = 2 \hbar v_F \sqrt{\pi n_c} $.  
Whereas tilted Dirac cones in borophene provide a distinct anisotropic optical
conductivity which can be seen in the subsequent discussion.
\begin{figure}[htbp]
\begin{center}\leavevmode
\includegraphics[width=85mm,height=85mm]{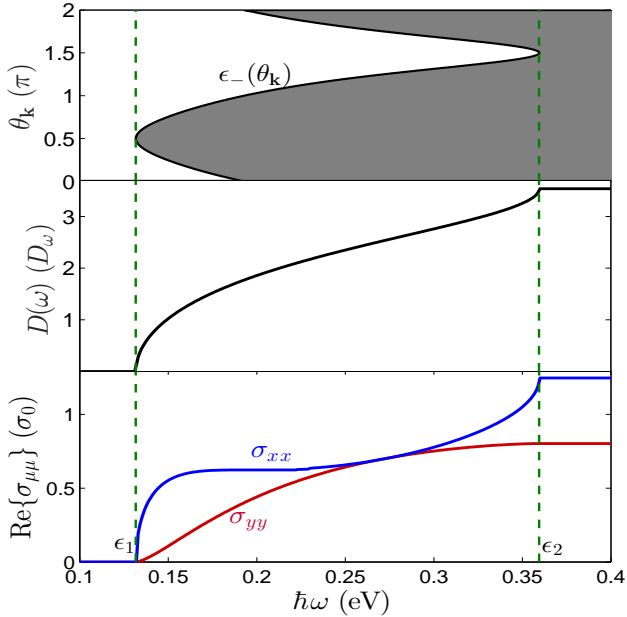}
\caption{Top panel: Plots of $\epsilon_{-}(\theta_{\bf k}) $ versus $\theta_{\bf k} $.
Middle panel: plots of the joint density of states versus photon energy $\hbar \omega $. 
Bottom panel: Plots of the optical conductivities ${\rm Re} \; \sigma_{xx} (\omega)$ 
and  $ {\rm Re} \; \sigma_{yy}(\omega)$ in units of $\sigma_0 = e^2/4\hbar$ versus 
photon energy $\hbar \omega $.
}
\label{Fig1}
\end{center}
\end{figure}
We analyze the real part of the optical conductivity by solving Eq. (\ref{opt1}) numerically
for electron density $ n_c = 1.0 \times 10^{16} $ m${}^{-2}$ at $T=0$. 
The plots of ${\rm Re} \; \sigma_{xx}(\omega)$ and ${\rm Re}\; \sigma_{yy}(\omega)$ 
as a function of photon energy $\hbar \omega $ are shown in the lower panel of Fig. 2.
It exhibits anisotropic nature of the optical conductivity. 
We plot $ \epsilon_{-}(\theta_{\bf k}) = 2 \hbar k_F^{-}(\theta_{\bf k}) \Delta (\theta_{\bf k}) $
in the top panel of Fig. 2. The shaded region in the top panel contributes to the 
optical conductivity.  
The optical transition from the valence band to the conduction band takes place when
the photon energy satisfies the inequality $ \hbar \omega \geq \epsilon_{-}(\theta_{\bf k})$.
The optical transition begins at $ \hbar \omega = 0.113 $ eV,
which corresponds to 
$ \epsilon_{1} = \epsilon_{-}(\pi/2) =  2E_F \frac{v_y}{v_y + v_t} < 2 E_F$. 
Moreover, the optical conductivities attain a maximum value when 
$\hbar \omega = 0.359 $ eV, which corresponds to
$\epsilon_2 = \epsilon_{-}(3\pi/2) = 2E_F\frac{v_y}{v_y - v_t} > 2E_F$.
Note that the two energy scales $\epsilon_1 $ and $ \epsilon_2$ depend on the
carrier density, tilted parameter $v_t$ and the velocity $v_y$ along the tilted direction.
We have checked numerically that 
$ {\rm Re} \; \sigma_{xx}(\omega) =  {\rm Re} \;  \sigma_{yy}(\omega) $ when 
$ \hbar \omega  \simeq 2 E_F $.
By knowing the energies $\epsilon_1 $ and $\epsilon_2$ from an experimental measurement,
one can extract the tilted parameter $v_t$ using the relation 
\begin{equation}
 v_t  = v_y E_F \Big[ \frac{1}{\epsilon_1} - \frac{1}{\epsilon_2} \Big].
\end{equation}
Analyzing the lower panel of Fig. 2, the maximum attainable absorptive part of the 
conductivity along the tilted direction is $\sigma_{yy}^{\rm max} =  \sigma_0 (v_y/v_x) < \sigma_0 $ 
and its orthogonal axis is $\sigma_{xx}^{\rm max} =  \sigma_0 (v_x/v_y) > \sigma_0 $.
It is interesting to note that $\sigma_{xx}^{\rm max} $ and $\sigma_{yy}^{\rm max}$ 
do not depend on the carrier density as well as the tilted parameter $v_t$.
Moreover, $ \sigma_{xx}^{\rm max} > \sigma_{yy}^{\rm max} $ and the product of these two conductivities 
$\sigma_{xx}^{\rm max} \cdot \sigma_{yy}^{\rm max} = \Big(\frac{e^2}{4\hbar}\Big)^2$ is universal.

To confirm the results of the optical conductivity, we 
analyze the joint density of states which is given by
\begin{eqnarray}
D(\omega) = \int_0^{2\pi} 
\frac{dC [f(E_-(k_\omega,\theta_{\bf k})) - f(E_+(k_\omega,\theta_{\bf k})) ]}
{4 \pi^2 |\partial_k E_g({\bf k})|_{E_g=\hbar \omega} },
\end{eqnarray}
where $C$ is the line element along the contour.
In the middle panel of Fig. 2, we show the joint density of states versus the photon
energy $\hbar \omega$.
One can easily see that that the van Hove singular points 
are at $\theta_{\bf k} = \theta_s = \pi/2,3\pi/2$.
The region of zero optical conductivity is nicely captured by the joint density
of states.  
\begin{figure}[htbp]
\begin{center}\leavevmode
\includegraphics[width=65mm,height=65mm]{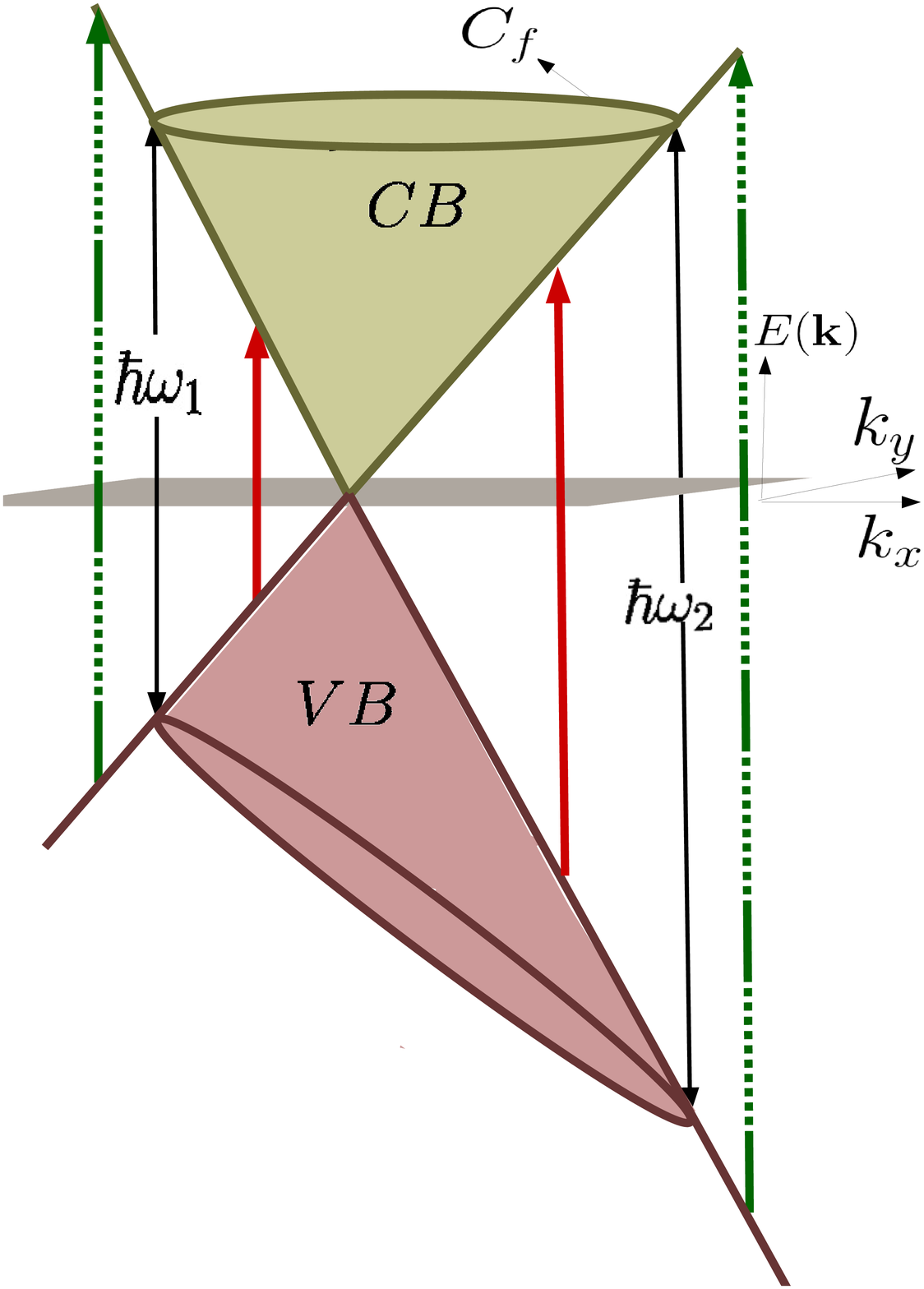}
\caption{Sketch of the allowed (dotted-dash-green) and forbidden (solid-red) inter-band 
transitions for $n$-doped borophene. 
}
\label{Fig1}
\end{center}
\end{figure}

The absorptive part of the optical conductivity arises due to the transitions 
from the valence band to the conduction band for a given momentum as demonstrated 
in Fig. 3. 
The green (dashed) and red (solid) arrows indicate the allowed and forbidden transitions,
respectively. 
One can easily see from this sketch that there are no allowed transitions 
for the photon energy $\hbar \omega < \epsilon_1$ as a result of the Pauli blocking. 
One can also notice that the transitions are allowed even for $\hbar \omega > \epsilon_2$. 


\section{Summary and conclusions}
We have presented detailed theoretical studies of the Drude weight and optical conductivity 
of the 8-$Pmmn$ borophene. 
The exact analytical expressions of the Drude weight, components of the optical conductivity and
the onset energy needed for initiating the optical transitions are provided.
We also obtain an analytical expression for the photon energy required to attain
maximum optical conductivity.
We find that the Drude weight and the absorptive parts of the optical conductivity 
are strongly anisotropic as a result of the tilted Dirac cones.
The tilted parameter $v_t$ and the velocity components
($v_x,v_y$) can be extracted from experimental measurements. We have shown that
the product of the maximum values of the anisotropic conductivities 
is always universal.

\section{Acknowledgements}
We would like to thank Arijit Kundu and SK Firoz Islam for useful discussion.

\appendix

\section{Alternative derivation of the optical conductivity}
We provide an alternative derivation of the optical 
conductivity using Green's function method..
The Kubo formula for the optical conductivity 
is given by 
\begin{eqnarray} \label{kuboG}
{}& \sigma_{\mu\nu}(\omega)=i\frac{e^2}{\omega}\frac{1}{(2\pi)^2}
\int d^2k \nonumber\\&\times T\sum_n \textrm{Tr} \langle \hat{v}_\mu \hat{G}({\bf k},\omega_n) 
\hat{v}_\nu \hat{G}({\bf k},\omega_n+\omega_l) \rangle_{i\omega_n 
\rightarrow \omega + i\delta}.
\end{eqnarray}
Here, $T$ is the temperature and $n$ and $l$ are integers, where $\omega_l=(2l+1) \pi T$ 
and $\omega_n = 2n \pi T$ are the fermionic and bosonic Matsubara frequencies, respectively.

The Green's function of the Hamiltonian in Eq. (\ref{Hamil}) is given by
\begin{eqnarray}\label{green}
\hat{G}({\bf k},\omega) & = & \sum_\lambda \Big[ \sigma_0 - \frac{\lambda}{\Delta(\theta_{\bf k})}
\big(v_x\cos\theta_{\bf k} \; \sigma_x + v_y \sin\theta_{\bf k} \; \sigma_y\big) \Big] \nonumber\\
& \times & G_\lambda({\bf k},\omega),
\end{eqnarray}
where
$ G_\lambda({\bf k},\omega) = [i\hbar\omega + \mu - E_\lambda({\bf k})]^{-1} $.
Using this Green's function, the following trace is obtained as
\begin{widetext}
\begin{eqnarray}
& &\textrm{Tr}\langle\hat{v}_x\hat{G}({\bf k},\omega_n)\hat{v}_x 
\hat{G}({\bf k}, \omega_n + \omega_l)\rangle = \sum_{\lambda,\lambda^\prime}
\Big[\frac{v_x^2}{2}(1-\lambda\lambda^\prime)+\lambda\lambda^\prime
\frac{v_x^4\cos^2\theta_{\bf k} }{\Delta^2(\theta_{\bf k})}\Big]G_\lambda({\bf k},\omega_l)
G_{\lambda^{\prime}}({\bf k},\omega_l+\omega_n).
\end{eqnarray}

Using the well-known identity
\begin{eqnarray}
&&T \sum_s \bigg[\frac{1}{i\hbar\omega_n+\mu-E_\lambda} \cdot 
\frac{1}{i\hbar(\omega_l+\omega_n)+\mu-E_{\lambda^\prime} }\bigg]= 
\begin{cases}
\frac{f(E_\lambda)-f(E_{\lambda^\prime})}{i\hbar\omega_n-E_{\lambda^\prime}+E_{\lambda}},& \text{if } 
\lambda\neq \lambda^\prime\\
    0,              & \text{otherwise.}
\end{cases}
\end{eqnarray}
One can further simply the above equation as
\begin{eqnarray}\label{trace}
&& T \sum_n \textrm{Tr}\langle \hat v_x \hat{G}({\bf k},\omega_n) \hat v_x 
\hat{G}({\bf k},\omega_l+\omega_n) \rangle = 
\frac{v_x^2v_y^2\sin^2\theta_{\bf k}}{\Delta^2(\theta_{\bf k})} 
\Big[\frac{f[E_-({\bf k})] - f[E_+({\bf k})]}{i\hbar\omega_n-E_+({\bf k}) + E_-({\bf k})}+ 
\frac{f[E_+({\bf k})]-f[E_-({\bf k})]}{i\hbar\omega_n-E_-({\bf k})+E_+({\bf k})}\Big].
\end{eqnarray}

It can be seen that the second term turns out to be zero as 
a result of the conservation of energy.
Using the result of Eq. (\ref{trace}) into Eq. (\ref{kuboG}), we have
\begin{eqnarray}
\sigma_{xx}(\omega)=\frac{i e^2}{(2\pi)^2\omega}\int_0^\infty\int_0^{2\pi}  dk d\theta_{\bf k} 
\frac{v_x^2v_y^2 \; k \sin^2\theta_{\bf k} }{\Delta^2(\theta_{\bf k})}
\frac{f[E_-({\bf k})]-f[E_+({\bf k})]}
{i\hbar\omega_n-E_+({\bf k})+E_-({\bf k})}\Big\vert_{i\omega_n\rightarrow\omega+i\delta}.
\end{eqnarray}
The real part of the optical conductivity is given by
\begin{eqnarray} \label{abc}
{\rm Re} \; \sigma_{xx}(\omega) =\frac{e^2}{4\pi}\int dk d\theta_{\bf k} 
\frac{v_x^2v_y^2\; k \sin^2\theta_{\bf k} }{\Delta^2(\theta_{\bf k})}
\Big[ f(E_-({\bf k})) - f(E_+({\bf k}))\Big] \delta(\hbar\omega - 2\hbar k\Delta(\theta_{\bf k})).
\end{eqnarray}
The above Eq. (\ref{abc}) can be further simplified to 
\begin{eqnarray} \label{sigxxA}
{\rm Re} \; \sigma_{xx}(\omega) =\frac{e^2}{16\pi}\int_0^{2\pi} d\theta_{\bf k}
\frac{v_x^2v_y^2\sin^2\theta_{\bf k} }{\Delta^4(\theta_{\bf k})} 
\Big[ f(E_-({k_\omega(\theta_{\bf k})})) - f(E_+ (k_\omega(\theta_{\bf k})))\Big]
\end{eqnarray}
\end{widetext}
where $k_\omega(\theta_{\bf k}) = \omega/2\Delta(\theta_{\bf k})$. 


Similarly, the $yy$ and $yx$ components of the optical conductivity can be obtained as
\begin{widetext}
\begin{eqnarray} \label{sigyyA}
{\rm Re} \; \sigma_{yy}(\omega) & = &  - \frac{e^2}{16\pi}\int_0^{2\pi} 
d\theta_{\bf k} \frac{v_x^2v_y^2\cos^2\theta_{\bf k} }{\Delta^4(\theta_{\bf k})} 
\Big[ f(E_+ (k_\omega(\theta_{\bf k}))) - f(E_- (k_\omega(\theta_{\bf k})))\Big], \\
{\rm Re} \; \sigma_{yx}(\omega) & = & \frac{e^2}{16\pi}\int_0^{2\pi} 
d\theta_{\bf k} \frac{v_x^2v_y^2\sin\theta_{\bf k} \cos\theta_{\bf k} }{\Delta^4(\theta_{\bf k})} 
\Big[f(E_+ (k_\omega(\theta_{\bf k}))) - f(E_- (k_\omega(\theta_{\bf k}))) \Big] \label{sigxyA}.
\end{eqnarray}
\end{widetext}

Equations (\ref{sigxxA}), (\ref{sigyyA}) and (\ref{sigxyA}) can be written in a 
compact form as given in Eq. (\ref{opt1}).

\end{document}